# Competitive Assessments for HAP Delivery of Mobile Services in Emerging Countries

Laurent Reynaud, Salim Zaïmi, Yvon Gourhant
Orange Labs
Lannion, France
{laurent.reynaud, salim.zaimi, yvon.gourhant}@orange-ftgroup.com

*Abstract*—In recent years, network deployment based on High Altitude Platforms (HAPs) has gained momentum through several initiatives where air vehicles and telecommunications payloads have been adapted and refined, resulting in more efficient and less expensive platforms. In this paper, we study HAP as an alternative or complementary fast-evolving technology to provide mobile services in rural areas of emerging countries, where business models need to be carefully tailored to the reality of their related markets. In these large areas with low user density, mobile services uptake is likely to be slowed by a service profitability which is in turn limited by a relatively low average revenue per user. Through three architectures enabling different business roles and using different terrestrial, HAP and satellite backhaul solutions, we devise how to use in an efficient and profitable fashion these multi-purpose aerial platforms, in complement to existing access and backhauling satellite or terrestrial technologies.

## I. INTRODUCTION

Connecting users from rural areas of emerging countries is a challenging task for service providers and for Mobile Network Operators (MNOs). In particular, reaching a rural population scattered on a large area requires for such a MNO a careful design of the intended financial model and related architecture. In essence, the business model must be profitable and the infrastructure must meet the basic Quality of Service (QoS) expected by the end users. In many large areas with low density of emerging countries, those natural requirements are hindered by multiple factors that include inexistent or undersized national backbones, wired networks and backhaul connectivity. Also, fluctuating and unreliable power sources impact negatively on the projected reliability and availability of the infrastructure. Those factors are further aggravated by the low projected Average Revenue Per User (ARPU) in the considered areas, which severely limits the MNO's network investment capacity, if its financial model is to be kept profitable. In order to address the aforementioned factors, we investigate the use of High Altitude Platforms (HAPs) in rural zones of emerging countries as a key component to deploy flexible and competitive network architectures.

HAPs [1], [2] are stratospheric stations, each composed of an aerial vehicle and a payload. They are either aerostats (i.e. lighter than air and able to keep a quasi stationary position) or aerodynes (i.e. airfoil based crafts) and generally operate between 17 km and 22 km. At such altitudes, aircrafts are well above the regular air traffic lanes and benefit from a local low of wind turbulence. Furthermore, there is almost no cloud formation and insulation can be maximized, which is an interesting aspect for solar powered crafts. HAP aircrafts are classified according to several categories that include manned planes, unmanned planes (whether solar or fuel powered) and unmanned aerostats. Vehicles from each type possess different advantages or constraints, for instance in terms of main (and, if applicable, secondary) power source, power and weight reserved for the payload, flight duration, form factor, mobility and technology maturity.

One important property of HAPs is their large coverage area. Depending on the antenna technology and the desired minimum elevation angles, a single HAP can cover areas of 30 to 300 km radius. This naturally contrasts with the majority of terrestrial wireless access networks, where base stations or access points must be significantly more densely scattered on the areas to cover. Even though HAPs are mostly still at a design stage rather than being commercially available, this technology is foreseen as a promising way to competitively deliver a wide range of end-user applications, including fixed or mobile services, temporary event or disaster relief support [1]. Moreover, it is also able to provide backhaul connectivity for remote networks where needed [2]. In order to anticipate the successful use of HAPs in complement or competition with existing access and backhauling satellite or terrestrial technologies, business models and related scenarios must be carefully elaborated and analyzed.

The rest of the paper is organized as follows. We first present in Section II several works related to the design and assessment of general HAP-based business models and we express the potential transposition issues that are likely to arise in the context of rural areas of emerging countries. In Section III, we introduce the main scenario and describe the requirements and constraints faced in an emerging country by a MNO that seeks to expand its network from an urban to a rural zone. In the context of this scenario, we then investigate and assess the profitability of three architectures, with different backhaul solutions: satellite-based, HAP-based, and supported

This work was partially supported by the ANR RESCUE project, grant ANR-10-VERS-003 of the French Agence Nationale de la Recherche.

by an integrated terrestrial-aerial-satellite network. Afterwards, we discuss in Section IV the main challenges related to the design and deployment in rural zones of a multi-role HAP-based architecture. We finally present the concluding remarks and the ongoing work.

## II. RELATED WORKS

In the CAPANINA project [2], the authors presented business models [4] adapted to a series of 6 scenarios (WLAN for trains, 3G and IEEE 802.11/802.16 backhaul connectivity, fixed broadband, broadcast/multicast delivery, temporary events and disaster relief servicing, 3G access). Two interacting roles were proposed: the HAP operator, which is in charge of deploying and maintaining the aerial platform (including the payload and network connectivity), and the service provider, which operates the ground segments and deploys end-user services. This work investigates the business models associated with these scenarios, from either HAP operator or service provider perspective. Capital and Operational Expenditures (OPEX and CAPEX), Net Present Value (NPV) and Internal Rate of Return (IRR), among other assessment metrics, are evaluated as far as the maturity of the HAP platforms allows. However, those use cases were tailored for western European countries and ARPU is that of established economies (e.g. end user broadband revenue of 40 € and broadcast/multicast revenue of 25 €), not of emerging markets. Also, the initial estimation of users (per HAP or per cell) and user growth rates apply for mature markets and cannot directly be adapted to the emerging countries environment. Furthermore, most proposed business models are particularly sensitive to both the customer prices and the number of users, which aggravates the aforementioned market transposition issues.

In [5], the authors sought to build a comprehensive collection of roles (including HAP operators, HAP vendors, service providers, mobile and fixed network operators and intermediate agents) in order to obtain a realistic, neutral and open ecosystem. However, the relations between HAP operators and Mobile Network Operators (MNOs) are insufficiently described: MNOs are systematically considered as resource consumers, while backhaul links are allegedly provided by fixed network operators. However, as MNOs need to interconnect their access equipments through backhaul links, they may be able to provide in turn backhaul connectivity to HAP operators, depending on the considered network topology. That is especially true in emerging countries, where fixed network infrastructures are often limited (if not non existent), especially in suburban and rural areas. Another problem with the transposition of the model developed in [5] into the emerging countries context is the imprecision of the related risk assessments.

The work presented in [3] is similar to [4] in terms of interactive roles (HAP operator and service provider). However, in order to evaluate the profitability of a 4G service over the South Korean peninsula, the model elaborated in [3] is based on the progressive cost decrease of further HAP units (according to the learning effect ) as the total number of HAP units produced increases. Moreover, it uses a diffusion-substitution model that expresses the transition process from 3G to 4G. For the economic analysis, the authors chose an ARPU of US 35 $, adapted to the South Korean market but which would be unrealistic for emerging countries. Also, the diffusion-substitution model does not take into account the multi-access nature of 4G, where HAP is only one access technology among others (e.g. terrestrial 4G base stations or access points) in a broader, and highly competitive, ecosystem.

## III. MAIN SCENARIO AND PROPOSED ARCHITECTURES

In this section, we describe a scenario where a MNO, which was so far operating a mobile network exclusively in urban zones of an emerging country, seeks to increase its customer base by expanding its network to a rural zone. It is presupposed that the national fiber backbone, if it exists, is too scarcely deployed to be able to interconnect each rural site without deploying a dedicated backhaul. Moreover, deploying such a full terrestrial backhaul network that connects each site (e.g. via fiber or microwave transmissions) is opted out because of deployment costs. For such a MNO, which already has a core network and an access network so far located in high density areas, the main difficulty is to deploy at a reasonable cost a network that is able to connect a rural population scattered on a large area. In the rest of this section, we will investigate and assess the profitability of three architectures, with different backhaul solutions: satellite-based, HAP-based, and supported by an integrated terrestrial-aerial-satellite network. We also make the preliminary assumptions described in Table I. With the given surface to cover and penetration rate, 18000 subscribers are expected. Also, 108 cell sites, each served by one Node B, must be interconnected to the existing MNO network via backhaul links. In the rest of this section, we describe two satellite- and HAP-based backhaul solutions.

### A. Satellite-based backhaul architecture

In this first architecture, we assess the costs required to provide a full satellite-based backhaul interconnection between the cell sites located in a rural zone of Uganda, and one remote aggregation site. The general topology is illustrated in Fig. 1. In our example, for each cell site, a link budget was estimated on the basis of several commercial geostationary satellites operating in Uganda in the C band [9] and able to provide uplink (UL) and downlink (DL) transmissions of respectively 2 Mb/s and 5 Mb/s on 72 MHz transponders, with a link availability of 99.96%. We therefore determined that a total bandwidth of 385 MHz was required on the space segment to support the service. We dimensioned accordingly the dedicated ground segment equipment: antennas with a diameter of 3.80 m for the cell sites and 11 m for the aggregation site, amplifiers, modems and adapted frequency transposition equipment.

TABLE I. GENERAL SCENARIO PARAMETERS

| | |
|---|---|
| Surface to cover (km$^2$) | 1800 |
| Penetration rate (subscribers per km$^2$) | 10 |
| 3GPP radio technology | High Speed Packet Access (HSPA) |
| Number of sites to interconnect | 108 |
| Total maximum throughput via the backhaul links (Mb/s) | 756 |

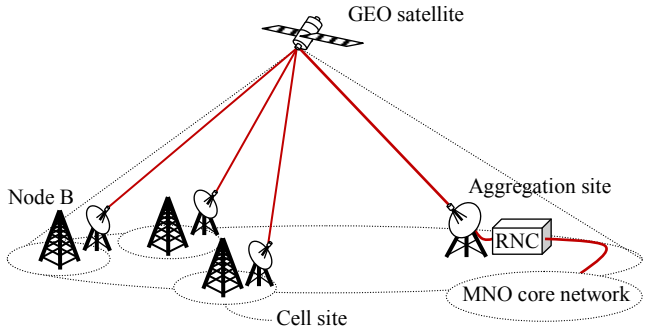

Figure 1. Architecture of a satellite backhaul solution interconnecting rural cell sites with a remote Radio Network Controler (RNC).

We then estimated the backhaul CAPEX and OPEX. The results are given in Table II, while the repartition in term of equipment is presented per cell and aggregation sites in Fig.2. For this architecture, the satellite ground equipment represents the totality of the cost. For the aggregation site, the antenna almost amounts for half of the backhauling costs. For the cell sites however, the cost is more equally divided between the antenna and the frequency transposition equipment, which each amount for about one third of the backhauling cost. Table II shows that with a cost of 920 k€, the aggregation site amounts for less than a sixth of the total 6.4 k€ CAPEX for the backhaul. Also, this architecture results in a high OPEX of 11.3 M€ (i.e. 105 k€ per cell site, including the related bandwidth on the aggregation site). This operational cost was estimated on the basis of an average commercial yearly lease price for 385 MHz of bandwidth on geostationary satellites operating in the C band in this region. Note that the estimated solution uses multiple channels per carrier (MCPC) [10] mechanisms, but no further improvement on the bandwidth usage was investigated; the use of optimized spectrum usage algorithms such as Carrier-in-Carrier (CnC) [11] would be likely to decrease the total OPEX. In any case, this architecture was mainly intended to provide a general reference to the HAP-based use cases, which themselves rely on relatively rough level of cost estimations.

*B. HAP-based backhaul architectures*

As previously mentioned, whether they belong to the manned / unmanned planes or unmanned aerostats categories, HAPs exhibit different constraints or advantages.

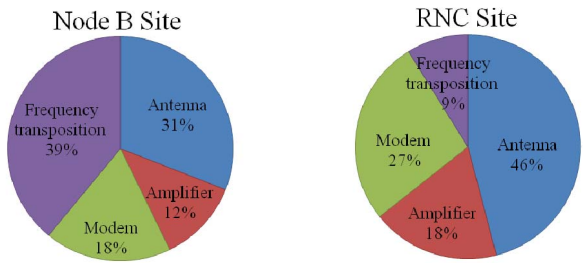

Figure 2. CAPEX generated by the backhaul equipment on cell (Node B) and aggregation (RNC) sites, for architecture 1.

TABLE II. CAPEX AND OPEX RESULTS FOR ARCHITECTURE 1

| | |
|---|---|
| CAPEX per cell site (k€) | 50 |
| CAPEX for the aggregation site (k€) | 920 |
| Total CAPEX for the backhaul (M€) | 6.4 |
| OPEX per cell site plus related fraction of OPEX on the aggregation site (k€) | 105 |
| Total OPEX (M€) | 11.3 |

This directly impacts the subsequent business models, in terms of capital expenditure (CAPEX) and operational expenditure (OPEX), or, on the contrary, in term of additional capacity and consequently, revenues. For instance, as [4] points out, while some stratospheric manned planes are already commercially available with a relatively mature technology and almost no HAP design costs, unmanned planes (whether solar or fuel powered) are still in development, and the associated models would require to be provisioned with additional cost and time. Models which rely on airships still require extensive developments for many core aspects (e.g. propulsion, stability, energy storage, hull design, inter-platform link availability…) and are exposed to maximal design costs and delays. On the contrary, airships (and manned airplanes) can support high capacity payloads (in terms of volume, weight and energy consumption), whereas unmanned airplanes offer payloads with significantly lower capacities.

We illustrate this tradeoff by using two types of HAPs in the following architectures. Note that for the sake of concision, only unmanned solar planes and airships, the operating and cost parameters of which are described in Table III, are considered in the scenario. However, compared to the listed operating characteristics, unmanned fuel planes performance would be close to that of unmanned solar planes, while manned airplane would exhibit similar characteristics to those described for unmanned airships. Those parameters are based on [4]. In particular, for both types of HAPs, the total throughput expressed in Table III represents a maximum fronthaul capacity, with a maximum clear air cell capacity of 120 Mb/s. While solar planes cover a similar area, they use a lower number of cells compared to the case of airships, and the average available throughput per surface is less than a fifth of that allowed by airships.

TABLE III. HAP OPERATING AND COST PARAMETERS

| | *Unmanned solar plane* | *Unmanned airship* |
|---|---|---|
| Number of available[a] cells | 18 | 97 |
| Covered area (km$^2$) | 2734 | 2827 |
| Total fronthaul throughput (Mb/s) in clear air | 2160 | 11640 |
| Average throughput per surface unit (Mb/s/km$^2$) | 0.20 | 1.03 |
| Gateway capacity (Mb/s) | 960 | |
| Development cost (M€) | 50 | 225 |
| CAPEX per HAP (M€) | 4 | 30 |
| OPEX per HAP (M€) | 1 | 4 |

a. Unmanned solar planes and airships are respectively based on a cellular layout of 19 cells with 5% redundancy and 121 cells with 20% redundancy [4].

However, estimated development costs, CAPEX and OPEX show that while airships are often seen as the best economical long term option [2], [3] thanks to their higher payload capacity, these types of HAPs are not necessarily adapted to every business model, because of their higher cost. In particular, a model where the required bandwidth per cell is expected to be relatively low, which is the case for instance in rural areas of emerging countries, is likely to benefit from the use of a lower capacity and less expensive HAP type such as an unmanned solar plane. In this context, those latter HAPs are thus more adapted to business models where mobile usages and subsequent bandwidth requirements slowly grow while a maximum coverage must be offered right from the start of the model projection. Note that we do not make particular assumptions about the usage of the HAP frequency spectrum. In this document, we simply state that the HAP fronthaul capacities given in Table III are compatible with the use of the 31/28 GHz millimeter band.

*1) Use case with the MNO as sole business role*

In this second architecture, we investigate the case where the MNO directly operates one or several HAPs in order to cover the desired area. Compared to the previous architecture, the satellite is replaced by the aerial platforms. In this case however, the MNO must support the costs of HAP development and acquisition. Also, in the previous architecture the aggregation site, thanks to the large satellite coverage may be located near the MNO core network. Here however, either the aggregation site must be totally within the covered zone, RNC included, or if the RNC is remotely located, the traffic must be backhauled to this equipment. And even in the case where the RNC is within the HAP coverage, this aggregation site must be interconnected to the MNO core network through a backhaul. So in either case, the MNO must also support the cost of a backhaul.

This architecture is illustrated in Fig. 3, where for this example, the aggregation site is split between a collection point located within the HAP coverage and is connected through a backhaul to a RNC remotely located within the MNO core network. We suppose that the backhaul link is partly deployed by the MNO and interconnects with a national backbone network, which in turn conveys the backhaul traffic to and from the RNC. This assumption, which is realistic in the case of Uganda, has an impact on the projected OPEX in terms of connection fees with this backbone. Also, in this architecture, on each cell site, the costly antenna and related equipment of the first architecture are replaced by less expensive industrial grade customer premises equipments (CPE). However, the collection point must support a gateway link with a capacity of 756 Mb/s and therefore requires a dedicated antenna and equipment. Regarding the type of HAP (either unmanned solar plane or airship), both types of platforms can serve the demanded coverage of 1800 km$^2$. They also both support the links to the cell sites (that represent, from the HAP, 756 Mb/s of fronthaul bandwidth) and they support the corresponding gateway link to the collection point, as previously mentioned. A point yet worth highlighting is that the total throughput given in Table III is only achieved in clear air conditions.

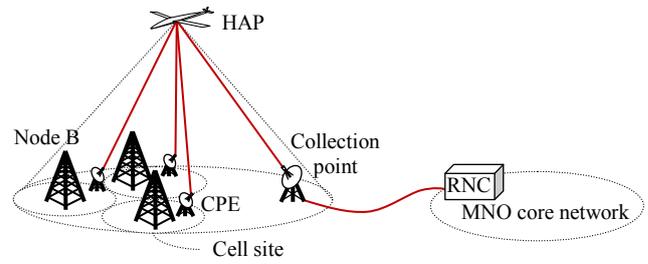

Figure 3. Architecture of a HAP backhaul solution interconnected to the RNC in the MNO core network through a terrestrial link.

When unmanned solar planes and airships need to support a link availability of 99.90%, the total fronthaul throughputs are decreased respectively to 360 Mb/s and 1940 Mb/s [2]. In this case, only an airship or a set of several interconnected planes could support the demanded availability. However, an unmanned solar plane could support the demanded traffic with an availability of 99% [2].

In our example, we will thus subdivide architecture 2 in case 2a, featuring an unmanned solar plane with fronthaul availability of 99%, and case 2b, that uses an unmanned solar airship with fronthaul availability of 99.90%. The OPEX and CAPEX results are given in Table IV for both cases, while the repartition in terms of different expenditures is presented for case 2a in Fig. 4. The repartition for case 2b, which is not represented, is similar to that of case 2a, with an even greater weight for HAP expenditures in both CAPEX and OPEX. Compared to the results related to architecture 1, cases 2a and 2b require less OPEX (with 6.4 M€ compared to respectively 1.1 M€ and 4.1 M€). Case 2b induces a greater CAPEX compared to architecture 1, with 30.9 M€ compared to 6.4 M€. But independently of any development cost considerations and taking only the lower OPEX into account, the related business model would still be likely to be more profitable over a 5- to 10-year projection period. In any case, case 2a is the most favorable both in terms of OPEX and CAPEX. However, as previously mentioned, this case offers the lowest fronthaul link availability.

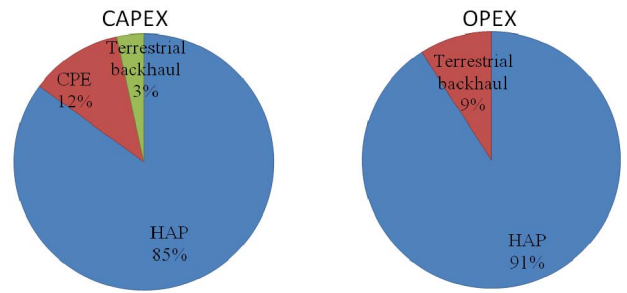

Figure 4. CAPEX and OPEX generated by the backhaul equipment for architecture 2a.

TABLE IV. CAPEX AND OPEX RESULTS FOR ARCHITECTURE 2

|  | 2a – unmanned plane | 2b – unmanned airship |
|---|---|---|
| Total CAPEX for the backhaul (M€) | 4.7 | 30.9 |
| Total OPEX for the backhaul (M€) | 1.1 | 4.1 |

Moreover, even without considering the development costs related to case 2a, the cost per user per month (with a network investment amortized over a 5-year period) would be about 9.5 €, which is still almost three times a projected 3.5 € ARPU.

*2) Use case with multiple business roles*

In this third architecture, we investigate the case of an integrated terrestrial-aerial-satellite architecture which enables three different business roles and related models. In this architecture, we consider the roles of the HAP operator, which, as previously mentioned, is in charge of deploying and maintaining one or several HAPs, and the service provider, which in this case is a MNO operating a mobile network in the considered area. Also, a third actor is the satellite operator; this latter role is not particularly discussed in this architecture because the related business model need not be adapted to the context of rural zones of emerging countries. As illustrated by Fig. 5, this architecture combines terrestrial, aerial and satellite networks. In particular, it integrates 3 types of backhaul links: as already seen for architecture 2, the MNO may use its own terrestrial backhaul. The HAP operator may also propose a terrestrial backhaul link (either its own, or otherwise made available through leasing or peering agreements with the backhaul owner). One or several HAPs may also use satellite backhaul links: the HAP operator may lease bandwidth to the satellite operator, and resell this capacity as permanent or temporary backhaul links to its own customers.

*a) HAP operator model*

In terms of expenditures, the situation of the HAP operator has similarities with what was seen in architecture 2 for the MNO: the deployment of the aerial equipment is initially costly, which likewise decreases the odds of profitability for the related business model. However, the situation of the HAP operator differs regarding its customer base: the main concern of this actor is not so much the end-user density of the considered area as the number of service providers which are in competition in this area and that could potentially hire some backhaul capacity. The eventual profitability of the discussed financial model hence depends on the openness and neutrality of the architecture, so that a maximum number of commercial partnerships can be established between the HAP operator and the MNOs present in the area. Moreover, flexibility is another key property of this architecture: the HAP operator must be able to initially deploy as few aerial platforms as possible, and to incrementally extend the existing infrastructure if additional capacity is required latter.

Another important requirement is the ability, for the HAP operator, to provide value-added services to MNOs. More particularly, in the discussed context where operators tend to deploy light and relatively inexpensive networks (compared to full 3GPP mobile architectures with costly backup mechanisms deployed on more mature markets), HAPs can provide valuable functionalities in terms of redundancy and contextual capacity increase for backhauling. In this regard, architecture 3 offers multiple possibilities to backhaul the traffic in a cell site. As an example, we propose a demand and income forecast in Table V as a part of a business model for the HAP operator.

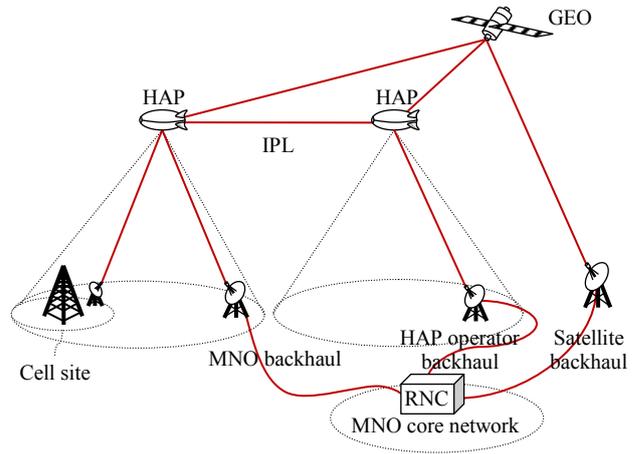

Figure 5. Architecture of an integrated terrestrial-aerial-satellite network. According to the deployment choices, the traffic from the cell sites may be backhauled in three different ways: through the MNO's own terrestrial backhaul, the HAP operator's terrestrial backhaul, or a satellite link. The architecture may use Inter-Platform Links (IPL) to backhaul the traffic.

Here, we assume the HAP operator is planning to deploy an unmanned solar plane, with 18 cells configured to offer links at 80 Mb/s with 99% availability and an illustrative contention ratio of 4:1. Three types of backhaul are proposed: a simple aerial backhaul link (with an indicative monthly price of 250 €), supposing the MNO possesses its own aggregation site connected to a backhaul. The second type of link (1000 € per month) is the same as in the first offer, with traffic being conveyed through the HAP operator's own backhaul if the MNO's backhaul temporarily fails. As the HAP operator may possess different types of backhaul (e.g. terrestrial and satellite backhaul as illustrated in Fig. 5), this offer is a means to offer redundancy and increased availability, which is an added-value service. The third offer (2000 € per month) is a complete backhaul link through the HAP operator's own backhaul, with increased availability, compared to the first offer. In this example, the evolution of the link demand, platform capacity and income are presented in Table V over a 10-year period. It illustrates the HAP operator's capacity to offer its wholesale customers discounted prices without impairing its model profitability.

*b) Service provider model*

One major problem with architectures 1 and 2 is that the initial CAPEX does not scale well with the number of cell sites.

TABLE V. DEMAND AND INCOME FORECAST FOR ARCHITECTURE 3

| Year | 0 | 3 | 6 | 9 |
|---|---|---|---|---|
| Aerial backhaul link 10 Mb/s | 192 | 164 | 148 | 230 |
| Aerial backhaul link 10 Mb/s with high availability | 86 | 172 | 260 | 260 |
| Complete backhaul link 10 Mb/s with high availability | 86 | 86 | 86 | 86 |
| Platform capacity (%) | 63.3 | 73.5 | 85.8 | 100 |
| Income (ex VAT) (M€) | 3.7 | 4.6 | 5.6 | 5.9 |

This is either because of the costly equipment required on the cell site itself (architecture 1) or because of the cost of the aerial platform unit that can support a limited number of cell sites only (architecture 2). With architecture 3 however, the MNO supports lighter initial costs, and has more flexible choices in terms of link availability and redundancy. For instance, the same operator, which in the case of architecture 2 would need to pay 9.5 € per subscriber per month, would only pay 2.1 € per subscriber per month in the case of architecture 3, for 108 simple aerial backhaul links as illustrated in Table V. This cost would then be under a projected 3.5 € ARPU.

IV. CHALLENGES FOR A MULTI-ROLE HAP ARCHITECTURE

One major challenge affecting the successful adoption of HAPs and integrated aerial networks is the maturity of the platforms [2]. Besides still requiring extensive development costs, unmanned fuel and solar aircrafts have reached different levels of development. However, those challenges are not only technical, but also depend on regulatory (e.g. management and cost of the related radio spectrum, airspace insertion rules), social and economic factors. Moreover, securing a viable activity for each HAP deployment largely relies on the ability of the HAP operators to durably offer competitive services. Examples of value-added services offered to terrestrial operators are access or backhauling connectivity with differentiated QoS (based on configurable link availability or contention ratio), seamless integration with sensor networks, redundancy mechanisms and also air monitoring of terrestrial access and backhaul networks [6]. Naturally, link redundancy and temporary capacity increase on large areas require advanced mechanisms such as controlled mobility [7], large scale network partitioning and clustering techniques [12] and an extended coordination of the HAP network through Inter-Platform Links (IPL). In any case, it is important to secure the neutrality, modularity and openness of the considered aerial infrastructure. To do so, the underlying architecture must benefit from the concept of infrastructure sharing [8] to HAPs. Along with other technologies (e.g. 3GPP LTE Radio Access Network) for which mobile infrastructure resources can be partitioned between several MNOs, HAP resources sharing should be beneficial to HAP operators and MNOs, both in terms of CAPEX and OPEX reduction. Furthermore, such a strategy should provide the means for MNOs to support large cost-effective radio coverage without tackling with hard issues related to HAP specific regulations, as those problems should mainly be left to HAP operators. In return, the latter should find the means to secure longer term trade agreements with multiple parties, and thus lessen the economic risks related to each HAP unit.

V. CONCLUSION

In this paper, we described the requirements and constraints faced by a mobile network operator seeking to expand its network from urban to rural zones of an emerging country. We studied the use of High Altitude Platforms (HAPs) in this context as a key component to deploy flexible and competitive network infrastructures. We then described three architectures offering different backhaul mechanisms: satellite-based, HAP-based, and supported by an integrated terrestrial-aerial-satellite network. We assessed the related business profitability from the perspective of several business roles, including HAP operator and service provider. We devised that implementing a sound business model and securing a long term viable activity for each HAP deployment greatly depends on HAP operators' capacity to keep their offers durably attractive for MNOs in comparison to other technologies. This is especially important in rural zones of emerging countries. Moreover, when relevant, HAP technologies should be part of a competitive multi-technology architecture where they are best needed and efficient. Also, HAP operators must carefully design their HAP modular payloads so as to draw optimal revenue from high value-added services.

In the future, we intend to further estimate the performance of an integrated terrestrial-aerial-satellite infrastructure and the profitability of its related business models by introducing in this architecture low altitude platforms and advanced mobility patterns allowed by mechanisms such as predicted and controlled mobility.

REFERENCES

[1] A. Aragón-Zavala, J.L. Cuevas-Ruiz and J.A. Delgado-Penin, "High-Altitude Platforms for Wireless Communications," Ed. John Wiley and Sons, Ltd., United Kingdom, 2008.

[2] D. Grace and M. Mohorčič, "Broadband Communications via High-Altitude Platforms," Ed. John Wiley and Sons, Ltd., United Kingdoms, 2010.

[3] J. Kim, D. Lee, J. Ahn, D.-S. Ahn, and B.-J. Ku, "Is HAPS Viable for the Next-Generation Telecommunication Platform in Korea?," EURASIP Journal on wireless communications and networking, 2008.

[4] G. Long, D. Grace, P. Likitthanasate, J. Kováts, L. Juhász et Al., "Applications and Business Models for HAP Delivery," FP6 IST CAPANINA Deliverable, CAP-D26-WP10-BTPUB-01, 2007.

[5] Z. Yang and A. Mohammed, "On the Cost-Effective Wireless Broadband Service Delivery from High Altitude Platforms with an Economical Business Model Design," Proceedings of the IEEE 68th Vehicular Technology Conference (VTC 2008-Fall), 2008.

[6] G.M. Galvan-Tejada and V. H. Correa-Cid, "HAP-Based Monitoring of the Emissions Produced by Terrestrial Microwave Systems," 4th International Conference on Electrical and Electronics Engineering (ICEEE 2007), Mexico City, Mexico, 2007.

[7] J. Rao and S. Biswas, "Controlled Node Mobility: A Mechanism for Energy Load Distribution in Sensor Networks," IEEE Global Telecommunications Conference (GLOBECOM'06), San Francisco, USA, 2007.

[8] D.-E. Meddour, T. Rasheed and Y. Gourhant, "On the role of infrastructure sharing for mobile network operators in emerging markets," Elsevier Computer Networks Journal, 2011.

[9] D. Roddy, "Satellite Communications," Fourth Edition, McGraw-Hill, 2007.

[10] M. Mwanakatwe and A.A. Richard, "An efficient and cost-effective solution for rural thin-route digital SCPC/MCPC satellite communications networks," Proceedings of the 10th International Conference on Digital Satellite Communications, Brighton, UK, 1995.

[11] C. Agne, M.B. Cornell, M. Dale, R. Kearns and F. Lee, "Shared-spectrum bandwidth efficient satellite communications," Proceedings of the IEEE Military Communications Conference (MILCOM'10), San Jose, USA, 2010.

[12] T. Rasheed, L. Reynaud and K. Al Agha, "A stable clustering scheme for large scale mobile ad hoc networks," Proceedings of the IEEE/Sarnoff Symposium on Advances in Wired and Wireless Communication, Princeton, USA, 2005.